%% file: paper.tex
\def\JPCM{{\it J. Phys. Condens Matter}\ }
\def\PR{{\it Phys. Rev.}\ }
\input{macros.tex}

\documentclass[twocolumn,showpacs,preprintnumbers,amsmath,amssymb]{revtex4}

\usepackage{graphicx}
\usepackage{dcolumn}
\usepackage{bm}
\begin{document}

\title{Electronic structure of random binary alloys : an augmented space formulation
in reciprocal space}

\author{Kamal Krishna Saha}
\author{Abhijit Mookerjee}
\affiliation{ S. N. Bose National Centre for Basic Sciences. Block-JD, Sector-III, \\
Salt Lake City, Kolkata-700098, India.}

\begin{abstract}
{We present here a reciprocal space formulation of the Augmented space recursion (ASR)
which uses the lattice translation symmetry in the full augmented space to produce
configuration averaged quantities, such as spectral functions and complex band structures. 
Since the real space part is taken into account {\sl exactly} and there is no truncation 
of this in the recursion, the results are more accurate than recursions in real space.  
We have also described the Brillouin zone integration procedure to obtain the configuration 
averaged density of states. We apply the technique to Ni$_{50}$Pt$_{50}$ alloy 
in conjunction with the tight-binding linearized muffin-tin orbital basis. 
These developments in the theoretical basis were necessitated by our future application to
obtain optical conductivity in random systems.}
\end{abstract}

\date{\today}

\pacs{71.23.-k}
\maketitle
\parindent 0pt

\section{Introduction}

The augmented space recursion carried out in a minimal basis set representation 
of the tight-binding linearized muffin-tin orbitals method (TB-LMTO-ASR) has been proposed earlier
by us \cite{asr,asr2} as an interesting technique for incorporation of the effects of
configuration fluctuations about the mean-field (the coherent potential 
approximation or the CPA) for random substitutionally disordered alloys.
 This can be achieved without the usual problems of violation
of the Herglotz analytic properties \cite{her} of the approximated configuration averaged Green functions
for the Schr\"odinger equation for these random alloys. Earlier we had  used this technique to look at
short-ranged ordering in such systems \cite{sro, sro2} , as well as local lattice distortions caused by
size difference between the constituents of the alloy \cite{latdis}.

One of the dissatisfying features of the method, and this has to with the recursion part, is the
truncation of the continued fraction expansion of the Green function. Truncation in the
configuration space part of the problem can be handled easily. We truncate out only those configurations
which occur with low probability and  contribute to the tail of the continued fraction. It is the
truncation in real space on which we do not have a controllable handle. Any truncation in real space
means that our recursion has been carried out on a finite cluster and edge effects become important. 
Quantities which converge fast are integrals of the density of states multiplied by well-behaved
functions of energy.
 We can also estimate the errors committed by truncating at a particular step
\cite{error}. However, the errors in the density of states itself cannot be controlled. This is
because even a small perturbation (like truncation after a large number of recursive steps) has a
profound effect on the spectrum of the Hamiltonian (see Haydock \cite{vol35}). The problem of truncation has always been laid at the door of the recursion method.

Is it not possible to  modify the TB-LMTO-ASR in such a way that the truncation is carried out only in
 configuration space ? One way of reducing the gigantic rank of the Hamiltonian in a real-space-labelled basis, is to go over to reciprocal space. In the $\k$ labelled basis, for a basis involving only   $s$, $p$ and $d$ states, the operators in reciprocal space have rank 9. However, to do this
we require lattice translational symmetry. In a random binary alloy, for instance, this is not immediately possible. However, the full augmented space, which is  the direct product of the
real space spanned by the site labelled basis $\{\underline{R}_i\}$ and the configuration space 
spanned by the configurations  of the system, possesses translational as well as
point group  symmetries \cite{saha}. Configurations of a binary alloy
can be labelled by a binary sequence of 0 and 1 (or $\uparrow$ and $\downarrow$ if Ising models appeal
to you more) and uniquely described by the {\sl cardinality sequence} $\{{\cal C}\}$, i.e. the sequence  
of positions where we have a 1 or a $\downarrow$ state. We had shown earlier that in the subspace
spanned by the reference states $\{\emptyset\}$, in which the configuration average is described, we
have full lattice translation symmetry provided the disorder is homogeneous \cite{comment}. The same
statement would be true if there is short-ranged order or local lattice distortions, provided the
short-ranged order or local lattice distortions are probabilistically identical anywhere in the system.
A consequence of this is that probability densities are independent of the site label and the configuration averaged quantity :

\begin{widetext} 
\[ \sum_{R_i}\sum_{R_j}\ \exp\left\{\rule{0mm}{4mm} \imath( \k\cdot R_i -\k'\cdot R_j)\right\}
\ \ll G(R_i,R_j,z)\gg \ =\ G(\k,z)\ \delta(\k-\k')\]
\end{widetext}

Based on this, we had proposed a TB-LMTO-Recursion in the reciprocal augmented space \cite{kasr}. The
recursion, as we shall show subsequently, is entirely in configuration space for each $\k$ label.
The truncation is also in configuration space alone and leads to calculation of the configuration
averaged spectral densities. These spectral densities are not a bunch of delta functions, as in
the case of ordered systems, but the complex self-energies, in general both energy and $\k$
dependent,  shift the peaks as well as and broaden them :
leading to  fuzzy, complex band structures.  

Although our method allows us to carry our augmented space recursion in reciprocal space, for many physical problems we need to carry our integration over the Brilluoin zone.
For instance, to obtain the density of states or optical conductivity\cite{optic} : 
\begin{widetext}
\begin{equation*}
 \ll n(E)\gg\ =\ \int_{BZ}\frac{d^3\k}{8\pi^3} \ll A(\k,E)\gg \end{equation*} 
\[\ll \sigma(\omega)\gg\ =\  \int\ dE\int_{BZ}\frac{d^3\k}{8\pi^3} \
{\mathrm Tr} \left[\rule{0mm}{4mm} {\mbf J}^{eff}(\k,E,\omega)
\ll {\mbf A}(\k,E)\gg {\mbf J}^{eff}(\k,E,\omega)^\dagger\ll{\mbf A}(\k,E+\omega)\gg\right]
\]
\end{widetext}

Another contribution of this paper is to modify the tetrahedral method of Brilluoin zone integration, so that we may carry out a smimilar integration technique for integrands which are smoother than the highly singular spectral functions of the ordered systems.
The proposed  Brilluoin zone integration is closely related to that
of Jepsen \cite{andersen} or Lehmann \cite{Lehmann}  for 
ordered systems.

\section{Augmented space Recursion in k-space}

The augmented space recursion based on the tight-binding linearized muffin-tin
orbitals method (TB-LMTO-ASR) has been described thoroughly in a series of articles \cite{kn:sdm,Am,KG,Am2,Am3,Am4}. 
We shall introduce the salient features of the ASR  which will be required by us in our
subsequent discussions.

 We shall start from a first principles tight-binding linearized muffin-tin 
orbitals (TB-LMTO) \cite{Ander1,Ander2} in the most-localized
representation ($\beta$ representation). This is necessary, because the subsequent
recursion requires a sparse representation of the Hamiltonian. In this $\beta$
representation, the second order alloy Hamiltonian is given by,

\[{\mathbf H}^{(2)} = {\mathbf E}_\nu + {\mathbf h} - \mathbf{hoh}  \]
where,
\begin{eqnarray}
{\mathbf h} \!\! &=& \!\!\!  \sum_{R} \left({\bf C}_{R} - {\mathbf E}_{\nu R}\right) {\cal P}_{R} 
 + \sum_{R}\sum_{R'}{\mathbf \Delta}_{R}^{1/2} \ {\mathbf S}_{RR'} \ {\mathbf \Delta}_{R'}^{1/2} \ {\cal T}_{RR'}\nonumber\\ 
\phantom{x}\nonumber\\
{\mathbf o} \!\! &=& \!\!\! \sum_R \mathbf{o}_R \ {\cal P}_R 
\end{eqnarray}

{\bf C}$_R$, {\bf E}$_{\nu R}$ , ${\mathbf\Delta}_R$  and ${\mathbf o}_R$ are diagonal matrices in angular momentum space :
\begin{eqnarray*}
&&{\mathbf C}_R = C_{RL} \ \delta_{LL'}\ \ ,\ {\mathbf E}_{\nu R}= E_{\nu RL} \ \delta_{LL'} \\
&& {\mathbf \Delta}_R = \Delta_{RL} \ \delta_{LL'} \ , \ {\mathbf o}_R=o_{RL} \ \delta_{LL'}
\end{eqnarray*}

\noindent and {\bf S}$_{RR'}$ is a matrix of rank $L_{max}$.
 ${\cal P}_{R}\ =\ \vert R\rangle\langle R\vert$ and ${\cal T}_{RR'}\ =\ 
\vert R\rangle\langle R'\vert$ are projection and transfer operators in the Hilbert
space ${\cal H}$ spanned by the tight-binding basis $\{|R\rangle\}$. Here, $R$ refers 
to the position of atoms in the solid  and $L$ is a composite label $\{\ell,m,m_s\}$ for 
the angular momentum quantum numbers. ${\bf C}$, ${\bf \Delta}$ and ${\bf o}$'s are potential 
parameters of the TB-LMTO method; ${\bf o}^{-1}$ has dimension of energy and ${\bf E}_\nu$'s are 
the energy windows about which the muffin-tin orbitals are linearized.

For a disordered binary alloy we may write~: 

\begin{eqnarray}
{ C}_{RL} & = &  C_L^A\  n_R + C_L^B\ \left( 1-n_R \right) \nonumber \\
\Delta_{RL}^{1/2} & = & \left( \Delta_L^A \right)^{1/2} n_R 
+ \left(\Delta_L^B \right)^{1/2}\left( 1-n_R \right)  \nonumber \\
{o}_{RL} & = &  o_L^A\  n_R + o_L^B\ \left( 1-n_R \right) 
\end{eqnarray}

Here $\{n_R\}$ are the random site-occupation variables which take values 1 and 0 
 depending upon whether the muffin-tin labelled by $R$ is occupied by $A$ or
$B$-type of atom. The atom sitting at $\{R\}$ can either be of type $A \ (n_R=1)$
with probability $x$ or $B \ (n_R=0)$ with probability $y$.
The augmented space
formalism (ASF) now introduces the space of configurations of the set of binary 
random variables  $\{n_R\}$ : $\Phi$. 

In the absence of short-ranged order, each random variable  $n_R$ has associated with it an operator  ${\bf M}_R$
whose spectral density is its probability density :

\begin{eqnarray}
p(n_R) &=& x \delta(n_R-1) + y\delta(n_R) \nonumber \\ 
&=& -\frac{1}{\pi} \lim_{\delta\rightarrow 0} \ \Im m 
\langle \uparrow_R|\left((n_R+i\delta) {\bf I}-{\bf M}_R\right)^{-1}|\uparrow_R\rangle \nonumber\\ 
\end{eqnarray}

where ${\bf M}_R$ is an operator whose eigenvalues $1, 0$ correspond to the observed values of $n_R$ and
whose corresponding eigenvectors $\{|1_R\rangle, |0_R\rangle\}$ span a configuration space
$\phi_R$ of rank 2.  We may change the basis to  $\{|\uparrow_R\rangle,|\downarrow_R\rangle\}$ (see
appendix A) and in these new basis the operator ${\bf M}_R$ is~: 


\begin{equation}
n_R \rightarrow {\bf M}_R = x{\cal P}^\uparrow_R + y{\cal P}^\downarrow_R + \sqrt{xy} \ ({\cal T}^{\uparrow\downarrow}_R 
+ {\cal T}^{\downarrow\uparrow}_R)
\label{asr1}
\end{equation}

Two new vectors span the space $\phi_R$. The full configuration space $\Phi$ = $\prod^\otimes_R\ \phi_R$ is then spanned by vectors of the form $\vert\uparrow\uparrow\downarrow\uparrow\downarrow\ldots\rangle$.
These configurations may be labelled by the sequence of sites $\{{\cal C}\}$ 
 at which we have a $\downarrow$. For example, for the state just quoted  $\{{\cal C}\}$
= $\vert\{3,5,\ldots\}\rangle$. This sequence is called the {\sl cardinality sequence}.
If we define the configuration $\vert\uparrow\uparrow\ldots\uparrow\ldots\rangle$ as the 
 {\sl reference} configuration, then the {\sl cardinality sequence} of the {\sl reference} configuration is the null sequence 
$\{\emptyset\}$.

The augmented space theorem \cite{Am} states that

\be 
\ll A(\{ n_{R}\}) \gg \eq < \{\emptyset\}\vert \widetilde{{\mbf A}}\vert \{\emptyset\}> 
\label{eq5}
\ee

\n where,

\[ \widetilde{\mbf A}(\{{{\bf M}_R}\}) \eq \int \ldots \int A(\{\lambda_{R}\})\ \prod d{\mbf P}(\lambda_{R}) \]

\n {\bf P}($\lambda_{R}$) is the spectral density of the self-adjoint operator ${{\bf M}}_{R}$. 

Applying this to (\ref{eq5}) to the Green function we get~:

\begin{equation}
\ll \o{G}(\k,z)\gg \ = \ \langle \k\otimes\{\emptyset\} |{(z{\bf\widetilde I} 
- {\bf\widetilde{H}}^{(2)})}^{-1} |\k\otimes\{\emptyset\} \rangle
\label{eq6}
\end{equation}

where {\bf G} and {\bf H}$^{(2)}$ are operators which are matrices in angular momentum space, and 
the augmented {\bf k-} space basis $|\k,L\otimes\{\emptyset\} \rangle$ has the form

\[ (1/\sqrt{N})\sum_R \mbox{exp}(-i\k\cdot R)|R\otimes \{\emptyset\}\rangle \]

The augmented space Hamiltonian ${\bf\widetilde H}^{(2)}$ is constructed from the TB-LMTO
Hamiltonian $\o{H}^{(2)}$ by replacing each random variable $n_R$ by  operators ${\bf M}_R$.
It is an operator in the augmented space 
$\Psi$ = ${\cal H} \otimes \Phi$. The ASF maps 
a disordered Hamiltonian described in a Hilbert space ${\cal H}$ onto an ordered Hamiltonian 
in an enlarged space $\Psi$, where the space $\Psi$ is constructed as the outer product of the
space ${\cal H}$ and configuration space $\Phi$ of the random variables of the disordered 
Hamiltonian. The  configuration space $\Phi$ is of rank 2$^{N}$ if there are $N$ muffin-tin 
spheres in the system. Another way of looking at ${\bf\widetilde H}^{(2)}$ is to note
that it is the {\sl collection} of all possible Hamiltonians for all possible
configurations of the system.

The resolvent of the Hamiltonian can be expressed in the following way~:

\begin{widetext}
\begin{eqnarray*}
{\left(z{\mathbf I}-{\mathbf H}^{(2)}\right)}^{-1} &=& {\left(z{\mathbf I}-{\mathbf C}-{\mathbf \Delta}^{1/2} \ {\mathbf S} \ {\mathbf \Delta}^{1/2} + \mathbf{hoh} \right)}^{-1} \\
&=& \mathbf{\Delta}^{-1/2}{\left[\frac{z\mathbf{I}-\mathbf{C}}{\mathbf\Delta}-\mathbf{S}+\left(\frac{\mathbf{C}-\mathbf{E_\nu}}{\mathbf\Delta}+\mathbf{S}\right)\left(\mathbf{\Delta}^{1/2}\mathbf{o}
\mathbf{\Delta}^{1/2}\right)\left(\frac{\mathbf{C}-\mathbf{E}_\nu}{\mathbf\Delta}+\mathbf{S}\right)\right]}^{-1}\mathbf{\Delta}^{-1/2}
\end{eqnarray*}
\end{widetext}

Expressions in bold are matrices in angular momentum space and other than {\bf S} and {\bf H}$^{(2)}$ and {\bf G} all others are diagonal matrices. 

In the above expression, since

\begin{eqnarray*}
\widetilde{\bf\Delta}^{-1/2}= \sum_R\left\{\rule{0mm}{5mm}{\bf A}({\bf\Delta}^{-1/2})\ {\cal P}_R\otimes{\cal I} 
+ {\bf B}({\bf\Delta}^{-1/2})\ {\cal P}_R\otimes{\cal P}^\downarrow_R \right.\dots \\
\left.\dots +{\bf F}({\bf\Delta}^{-1/2})\ {\cal P}_R\otimes\left({\cal T}^{\uparrow\downarrow}_{R}
+{\cal T}^{\downarrow\uparrow}_R\right)\right\}
\end{eqnarray*}

where for any diagonal (in angular momentum space) operator {\bf V} : 

\begin{eqnarray*}
&&{\mathbf A}({\mathbf V}) = A(V_L)\ \delta_{LL'}\qquad A(V_L) = x\ V^A_L + y\ V^B_L \\
&&{\mathbf B}({\mathbf V}) = B(V_L)\ \delta_{LL'}\qquad B(V_L) = (y-x)\ (V^A_L -  V^B_L) \\
&&{\mathbf F}({\mathbf V}) = F(V_L)\ \delta_{LL'}\qquad F(V_L) = \sqrt{xy}\ (V^A_L -  V^B_L) 
\end{eqnarray*}

we get :

\begin{eqnarray*}
\bf\widetilde{\Delta}^{-1/2}|\k\otimes\{\emptyset\}\rangle &=& \mathbf{A}(\mathbf{\Delta}^{-1/2})|\k\otimes\{\emptyset\} \rangle\dots\\ 
\dots&+&\mathbf{F}(\mathbf{\Delta}^{-1/2}) |\k\otimes\{R\}\rangle = |1\}
\end{eqnarray*}

The ket $|1\}$ is not normalized and we define the normalized ket as, 
$|1\rangle = [{\bf A}({\bf\Delta}^{-1})]^{-1/2} |1\}$. Then we may rewrite (\ref{eq6}) as 

\begin{eqnarray*}
\ll {\mathbf G(\k,z)}\gg &=& \langle 1\vert \left(z{\bf\widetilde I}-\widetilde {\mathbf A}+\widetilde {\mathbf B}+\bf\widetilde{F}-\bf\widetilde{ S}\right.\dots \\
&&\left.\dots+(\bf\widetilde{ J}+\bf\widetilde{S}) \ \bf\widetilde{ o} \ (\bf\widetilde{J}+\bf\widetilde{ S}) \right)^{-1} 
\vert 1 \rangle
\end{eqnarray*}

where, 

\begin{eqnarray}
\bf\widetilde{A} & = & \sum_{R} \left\{ \mathbf{A}(\mathbf{C}\mathbf{\Delta}^{-1})/\mathbf{A}(\mathbf{\Delta}^{-1}) \right \}\enskip {\cal P}_R\otimes {\cal I} \nonumber\\
\bf\widetilde{B} & = & \sum_{R} \left\{\mathbf{B}\left((z\mathbf{I}-\mathbf{C})\mathbf{\Delta}^{-1}\right)/\mathbf{A}(\mathbf{\Delta}^{-1}) \right\} \enskip 
{\cal P}_{R} \otimes {\cal P}^{\downarrow}_{R} \nonumber\\
\bf\widetilde{F} & = & \sum_{R} \left\{\mathbf{F}\left((z\mathbf{I}-\mathbf{C})\mathbf{\Delta}^{-1}\right)/\mathbf{A}(
\mathbf{\Delta}^{-1}) \right\}\enskip {\cal P}_{R}\otimes \left\{ {\cal T}_{R}^{\uparrow\downarrow} + {\cal
T}_{R}^{\downarrow\uparrow} \right\}  \nonumber\\
\label{eq7}
\end{eqnarray}

and $\widetilde {\bf J}=\widetilde{\bf  J}_A+\widetilde {\bf J}_B+\widetilde {\bf J}_F$ and $\widetilde{\bf o}=\widetilde {\bf o}_A+\widetilde {\bf o}_B+\widetilde {\bf o}_F$ where :

\begin{eqnarray}
\bf\widetilde{J}_A & = & \sum_{R} \left\{\o{A}((\o{C}-\o{E}_{\nu })\o{\Delta}^{-1})/\o{A}(\o{\Delta}^{-1})\right\}\enskip 
{\cal P}_R \otimes {\cal I} \nonumber \\
\bf\widetilde{J}_B & = & \sum_{R} \left\{\o{B}((\o{C}-\o{E}_{\nu })\o{\Delta}^{-1})/\o{A}(\o{\Delta}^{-1})\right\}\enskip 
{\cal P}_{R}\otimes {\cal P}^{\downarrow}_{R} \nonumber \\
\bf\widetilde{J}_F & = & \sum_{R} \left\{\o{F}((\o{C}-\o{E}_{\nu })\o{\Delta}^{-1})/\o{A}(\o{\Delta}^{-1})\right\}
\enskip {\cal P}_{R}\otimes \left\{ {\cal T}_{R}^{\uparrow\downarrow} + {\cal
T}_{R}^{\downarrow\uparrow} \right\}  \nonumber \\ 
\bf\widetilde{o}_A & = & \sum_{R} \{\o{A}(\o{o}\o{\Delta}) \ \o{A}(\o{\Delta}^{-1})\} \enskip 
{\cal P}_R \otimes {\cal I} \nonumber \\
\bf\widetilde{o}_B & = & \sum_{R} \{\o{B}(\o{o}\o{\Delta}) \ \o{ A}(\o{\Delta}^{-1})\} \enskip
{\cal P}_{R} \otimes {\cal P}^{\downarrow}_{R}\nonumber \\ 
\bf\widetilde{o}_F & = & \sum_{R} \{\o{F}(\o{o}\o{\Delta}) \ \o{A}(\o{\Delta}^{-1})\} 
\enskip {\cal P}_{R}\otimes \left\{ {\cal T}_{R}^{\uparrow\downarrow} + {\cal
T}_{R}^{\downarrow\uparrow} \right\}   
\label{eq8}
\end{eqnarray}
 
In case there is no off-diagonal disorder due to local lattice distortion because of size mismatch :

\[ {\bf\widetilde S} = \sum_{R}\sum_{R'} \o{A}(\o{\Delta}_R^{-1})^{-1/2}\ \o{S}_{RR'} \ \o{A}(\o{\Delta}_{R'}^{-1})^{-1/2} \enskip
{\cal T}_{RR'}\otimes {\cal I} \]

This equation is now exactly in the form in which recursion method may be applied. At this point
we note that the above expression for the averaged $G_{LL}(\k,z)$ is {\sl exact}.

The recursion method addresses inversions of infinite matrices \ref{}. Once a sparse representation
of an operator in Hilbert space, ${\bf\widetilde H}^{(2)}$, is known in a countable basis, the recursion method
obtains an alternative basis in which the operator becomes tridiagonal. This basis and the 
representations of the operator in it are found recursively through a three-term recurrence relation~:

\begin{equation}
|u_{n+1}\} = {\bf\widetilde H}^{(2)} |u_n\} - \alpha_n(\k) |u_n\} - \beta_n^2(\k) |u_{n-1}\}
\end{equation}

with the initial choice $|u_1\}=|RL\rangle\otimes|1\rangle$ and $\beta_1^2=1$. The recursion 
coefficients $\alpha_n$ and $\beta_n$ are real and are obtained by imposing the ortho-normalizability 
condition of the new basis set as~:

\begin{eqnarray*}
&&\alpha_n(\k) = \frac{\{n|{\bf\widetilde H}^{(2)}|n\}}{\{n|n\}} \phantom{x} ; \phantom{xx} \beta_{n-1}^2(\k) = \frac{\{n-1|{\bf\widetilde H}^{(2)}|n\}}{\{n|n\}} \phantom{x} \\
&& \mbox{and also} \phantom{xx} \{m|{\bf\widetilde H}^{(2)}|n\} = 0 \mbox{  for } m\not= n, n\pm 1
\end{eqnarray*}
 
To obtain the spectral function we first write the configuration averaged $L$-projected Green
functions as continued fractions~:

\begin{widetext}
\[ \ll G_{LL}(\k,z) \gg \ = \ \frac{{\beta_{1L}^2}}
        {\displaystyle z-\alpha_{1L}(\k)-\frac{\beta^2_{2L}(\k)}
        {\displaystyle z-\alpha_{2L}(\k)-\frac{\beta^2_{3L}(\k)}
        {\displaystyle \frac{\ddots}
        {\displaystyle z-\alpha_{NL}(\k)-\mathbf \Gamma_L(\k,z)}}}} \]
\end{widetext}

where $\mathbf \Gamma_L(\k,z)$ is the asymptotic part of the continued fraction.  
The approximation involved has to do with the termination of this continued fraction. The coefficients 
are calculated exactly up to a finite number of steps $\{\alpha_n,\beta_n\}$ for $n < N$ and the asymptotic
part of the continued fraction is obtained from the initial set of coefficients using the idea of Beer and
Pettifor terminator \cite{bp}. Haydock and coworkers \cite{kn:hay} have carried out extensive studies of 
the errors involved and precise estimates are available in the literature. Haydock \cite{kn:hay2} has shown
that if we carry out recursion exactly up to $N$ steps, the resulting continued fraction maintains the first 
$2N$ moments of the exact result.

It is important to note that the operators ${\bf\widetilde A}, {\bf\widetilde B}, {\bf\widetilde F}$ are all projection operators
in real space ({\sl i.e} unit operators in {\bf k-} space) and acts on an augmented space basis only to 
change the configuration part ({\sl i.e.} the cardinality sequence $\{{\cal C} \}$ ).

\begin{eqnarray*}
{\bf\widetilde A} \vert\vert \{ {\cal C} \} \rangle &=& A_{1} \vert\vert \{ {\cal C} \}\rangle \\
{\bf\widetilde B} \vert\vert \{ {\cal C} \} \rangle &=& A_{2} \vert\vert \{ {\cal C} \} \rangle \ 
\delta( R \in \{ {\cal C} \})\\
{\bf\widetilde F} \vert\vert \{ {\cal C} \} \rangle &=& A_{3} \vert\vert \{ {\cal C} \pm R \} \rangle
\end{eqnarray*}

The coefficients $A_{1} - A_{3}$ can been expressed from equation (\ref{eq7}). Similar expressions
hold for the operators in equation (\ref{eq8}). The remaining operator ${\bf\widetilde S}$ is 
diagonal in {\bf k-} space and acts on an augmented space only to change the configuration part~:

\begin{equation}
{\bf\widetilde S} \vert\vert \{ {\cal C} \} \rangle = \sum_{\chi} \exp{(-\imath {\bf k}.\chi) } \vert\vert \{ {\cal C} - \chi \} \rangle \nonumber
\end{equation}

Here $\chi$s are the near neighbour vectors. The operation of the effective Hamiltonian 
is thus entirely in the configuration space and the calculation does not involve the space 
${\cal H}$ at all. This is an enormous simplification over the standard augmented
space recursion described earlier \cite{kn:sdm,Am2,Am3,Am4}, where the entire reduced real
space part as well as the configuration part was involved in the recursion process. Earlier 
we had to resort to symmetry reduction of this enormous space in order to make the recursion 
tractable. Here the rank of only the configuration space is much smaller and we may
further reduce it by using the local symmetries of the configuration space, as described 
in our earlier letter \cite{kn:sdm}. However, this advantage is offset by the fact that the
effective Hamiltonian is energy dependent. This means that to obtain the  Green functions 
we have to carry out the recursion for each energy point. This process is simplified by 
carrying out recursion over a suitably chosen set of {\sl seed energies} and interpolating 
the values of the coefficients across the band.

\section{Spectral Density and Band Energy}

The self-energy which arises because of scattering by the random potential fluctuations is
of the form :

\[ \Sigma_L(\k,z) \ = \ {\frac{\beta^2_{2L}(\k)}
        {\displaystyle z-\alpha_{2L}(\k)-\frac{\beta^2_{3L}(\k)}
        {\displaystyle \frac{\ddots}
        {\displaystyle z-\alpha_{NL}(\k)-\mathbf \Gamma_L(\k,z)}}}}         \]

So the continued fraction can be written in the form $1/{(z-\tilde E_L(\k)-\Sigma_L(\k,E))}$, 
where $\tilde E_L(\k)=\alpha_{1L}(\k)$. 

The average spectral function $\ll A_\k(E) \gg$ is related to the averaged Green function in 
reciprocal space as :

\[ \ll A_\k(E) \gg \ = \sum_L \ll A_{\k L}(E) \gg \]
where,
\[ \ll A_{\k L}(E) \gg \ = -\frac{1}{\pi} \lim_{\delta \rightarrow 0+} \{\Im m \ll G_{LL}(\k,E-i\delta)\gg\} \]

To obtain the complex bands for the alloy we fix a value for $\k$ and solve for~:
\[ z-\tilde E_L(\k)-\Sigma_L(\k,E) = 0 \]

The real part of the roots will give the position of the bands, while the imaginary part of
roots will be proportional to the lifetime. Since the alloy is random, the bands always have
finite lifetimes and are fuzzy.

\section{Integration in {\bf k-} space}

To obtain the density of states we  need to integrate over the Brillouin zone 

\begin{eqnarray}
\ll n(E) \gg & = & \sum_L \sumk \ll A_{\k L}(E) \gg 
\label{sp}
\end{eqnarray}
 
For ordered systems the spectral function is  a bunch of delta functions :
$A^0_\k(E)\ =\ \sum_j\ A_j\delta(E-E_j(\k))$, with $j$ labeling a particular 
energy band. The integrand being highly singular, the integral (\ref{sp})
has to be calculated carefully. Jepsen {\em et. al.} \cite{andersen} and Lehmann 
\cite{Lehmann} had proposed an accurate technique : the tetrahedron method, for
obtaining such integrals accurately. In this section we shall discuss an extension
of that method for application to disordered systems.

In the presence of disorder the spectral function is smoother and we may rewrite
it in terms of the real and imaginary parts of the disorder induced self-energy :

\begin{equation}
\ll A_{\k L}(E) \gg 
\ = \ \frac{-\Sigma_L^I(\k,E)/\pi}{{\left(E-\tilde E_L(\k)-\Sigma_L^R(\k,E)\right)}^2
+\Sigma_L^I(\k,E)^2} \\
\end{equation}

Such a function is peaked around the zeroes of $E-\tilde{E}_L(\k)-\Sigma_L^R(\k,E)$ and the
$\Sigma_L^I(\k,E)$ provides the width of the peaks. The spectral function behaves roughly
as {\sl Lorentzian} in the vicinity of its peaks. We may reduce the above
expression to one amenable to the tetrahedron integration form by the following trick :
\begin{eqnarray*}
&=&\!\!\! \int dE' \frac{-\Sigma_L^I(\k,E)/\pi}{{\left(E-E'-\Sigma_L^R(\k,E)\right)}^2
+\Sigma_L^I(\k,E)^2} 
\delta\left(E'-\tilde E_L(\k)\right)\  \\
&=&\!\!\! \int dE' \ {\cal W}_{\k L}(E,E') \ \delta\left(E'-\tilde E_L(\k)\right)
\end{eqnarray*}

where ${\cal W}_{\k L}$ is defined as a {\sl weight function}. 
Now integrating above over the Brillouin zone, we may get configuration averaged density of states (DOS)~:

\begin{eqnarray*}
 \ll n(E) \gg &=& \sum_L \sumk \int dE' \ll A_{\k L}(E) \gg \\
&=& \sum_L \int dE' \sumk {\cal W}_{\k L}(E,E') \ \delta\left(E'-\tilde E_L(\k)\right)
\end{eqnarray*}

At this stage, in order to simplify notation we shall drop the $L$ index from all $L$ dependent
factors and understood that the eventual result is summed over all $L$.
In order to perform the above integration over BZ, we have generalized {\sl tetrahedron method} developed by Jepsen 
{\emph et al} \cite{andersen} and Lehmann {\emph et al} \cite{Lehmann} to include the weight function 
${\cal W}_k(E,E')$. We have followed the idea of MacDonald {\emph et al} 
\cite{MacDonald}. In this generalization the energies as well as the weight functions are linearly 
interpolated throughout the vertices of small tetrahedrons. We label the energies at the vertices
of the $i$th tetrahedron $\tilde E_1^i,\tilde E_2^i,\tilde E_3^i \ \mbox{and} \ \tilde E_4^i$, where the
indices correspond to increasing magnitude of the energy, i.e., $\tilde E_1^i\ge\tilde E_2^i\ge\tilde E_3^i 
\ge\tilde E_4^i$ and the corner values of the weight function be ${\cal W}_1^i,{\cal W}_2^i,{\cal W}_3^i \ 
\mbox{and} \ {\cal W}_4^i$. Then the averaged DOS may be written as~:  

\begin{equation}
\ll n(E)\gg = V_{MZ}\ \int dE'\ \sum_{i=1}^N\  C^i \ \sum_{k=1}^4\  I_k^i \ {\cal W}_k^i 
\end{equation}

where $I_k^i=I_k(E,E',\tilde E_1^i,\tilde E_2^i,\tilde E_3^i,\tilde E_4^i)$, $N$ is the 
number of tetrahedral micro-zones and $V_{MZ}$ is the micro-zone volume and also, \\

for $\tilde E_1^i<E'\le \tilde E_2^i $ \\
\begin{eqnarray*}
C^i   &=& 3 \ F_{21}F_{31}F_{41}/(E'-\tilde E_1) \\
I_1^i &=& \left(F_{12}+F_{13}+F_{14}\right)/3 \\
I_k^i &=& F_{k1}/3,\phantom{x} k=2,3,4. \\ \\ 
\end{eqnarray*}
\vskip -1cm
for $\tilde E_2^i<E'\le \tilde E_3^i$ \\
\begin{eqnarray*}
C^i   &=& 3 \left(F_{23}F_{31}+F_{32}F_{24}\right)/ E_{41} \\
I_1^i &=& F_{14}/3+F_{13}F_{31}F_{23}/C^i E_{41} \\ 
I_2^i &=& F_{23}/3+F_{24}^2 F_{32}/C^i E_{41}   \\
I_3^i &=& F_{32}/3+F_{31}^2 F_{23}/C^i E_{41}   \\
I_4^i &=& F_{41}/3+F_{42} F_{24} F_{32}/C^i E_{41} \\ \\ 
\end{eqnarray*}
\vskip -1cm
for  $\tilde E_3^i<E'\le \tilde E_4^i$ \\ 
\begin{eqnarray*}
C^i   &=& 3 \ F_{14} F_{24} F_{34}/(\tilde E_4-E') \\
I_k^i &=& F_{k4}/3,\phantom{x} k=1,2,3  \\  
I_4^i &=& \left(F_{41}+F_{42}+F_{43}\right)/3
\end{eqnarray*}

where $E_{mn}=\tilde E_m-\tilde E_n$ and $F_{mn}=(E'-\tilde E_n)/E_{mn}$. \\
Also $\ll n(E)\gg$ is
zero for $E'\le \tilde E_1^i$ or $E'\ge \tilde E_4^i$.  

\begin{figure}[t]
\begin{center}
\includegraphics[width=3.2in, height=2.5in]{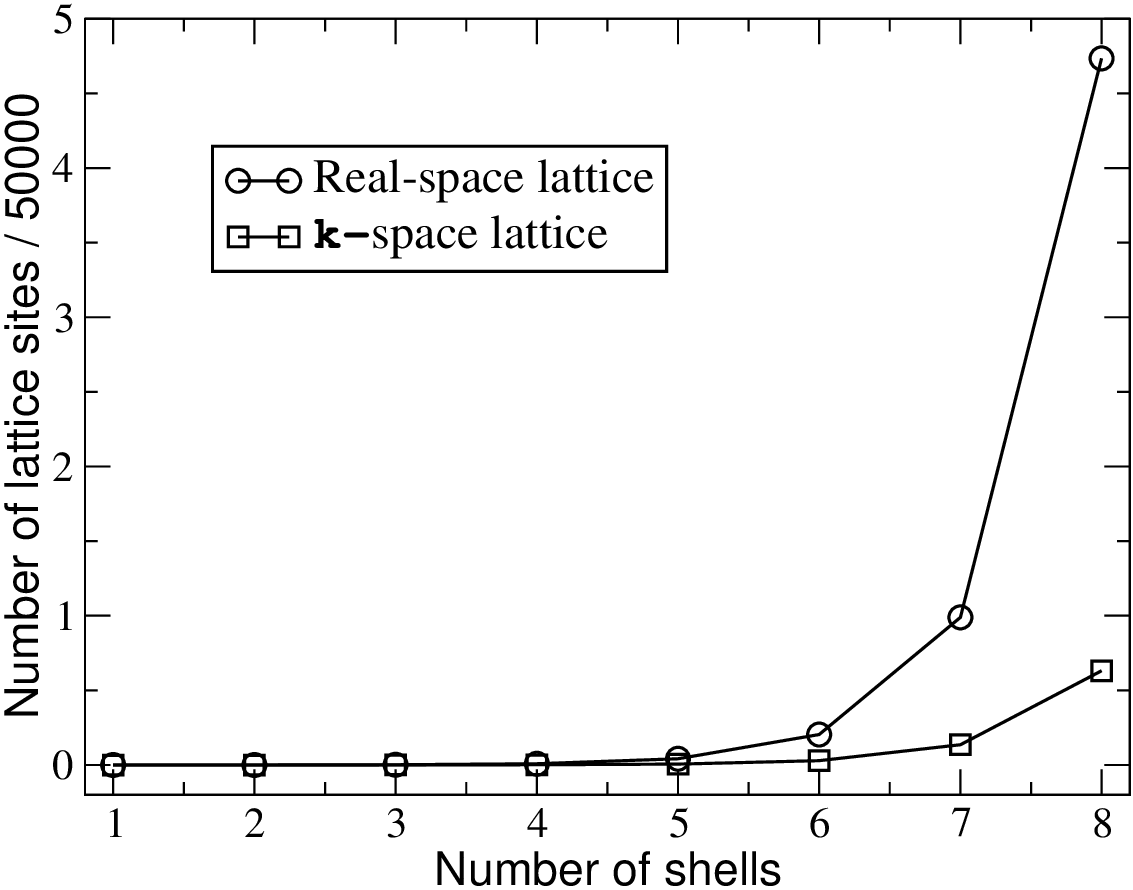}
\caption{\label {fig1}
Showing how the number of lattice sites increase with increasing the number
of shells in the real space and reciprocal space map. }
\end{center}
\end{figure}

\section{Computational details and Results}

For ordered faces the calculations have been performed in the basis of linear muffin-tin orbitals 
in the atomic sphere approximation including combined corrections. The scalar-relativistic 
calculations in this case are carried out for equal atomic spheres. The {\bf k-} space 
integration was carried out with $16\times 16\times 16$ mesh resulting 145 k-points 
for cubic primitive structure in the irreducible part of the Brillouin zone.

In Fig.~\ref{fig1} shows how the size of the augmented space map (in both {\bf k-} space and real-space
representation) increases as we increase the number of nearest neighbour shells from a starting site. 
We note that the reciprocal space map at a particular recursion step is much smaller than the
real augmented space map. This is because in the reciprocal augmented space we generate only
the different configurations. The full real space lattice map has been collapsed using
lattice translational symmetry in full augmented space. 

\begin{figure*}
\begin{center}
\includegraphics[width=6.0in, height=2.5in]{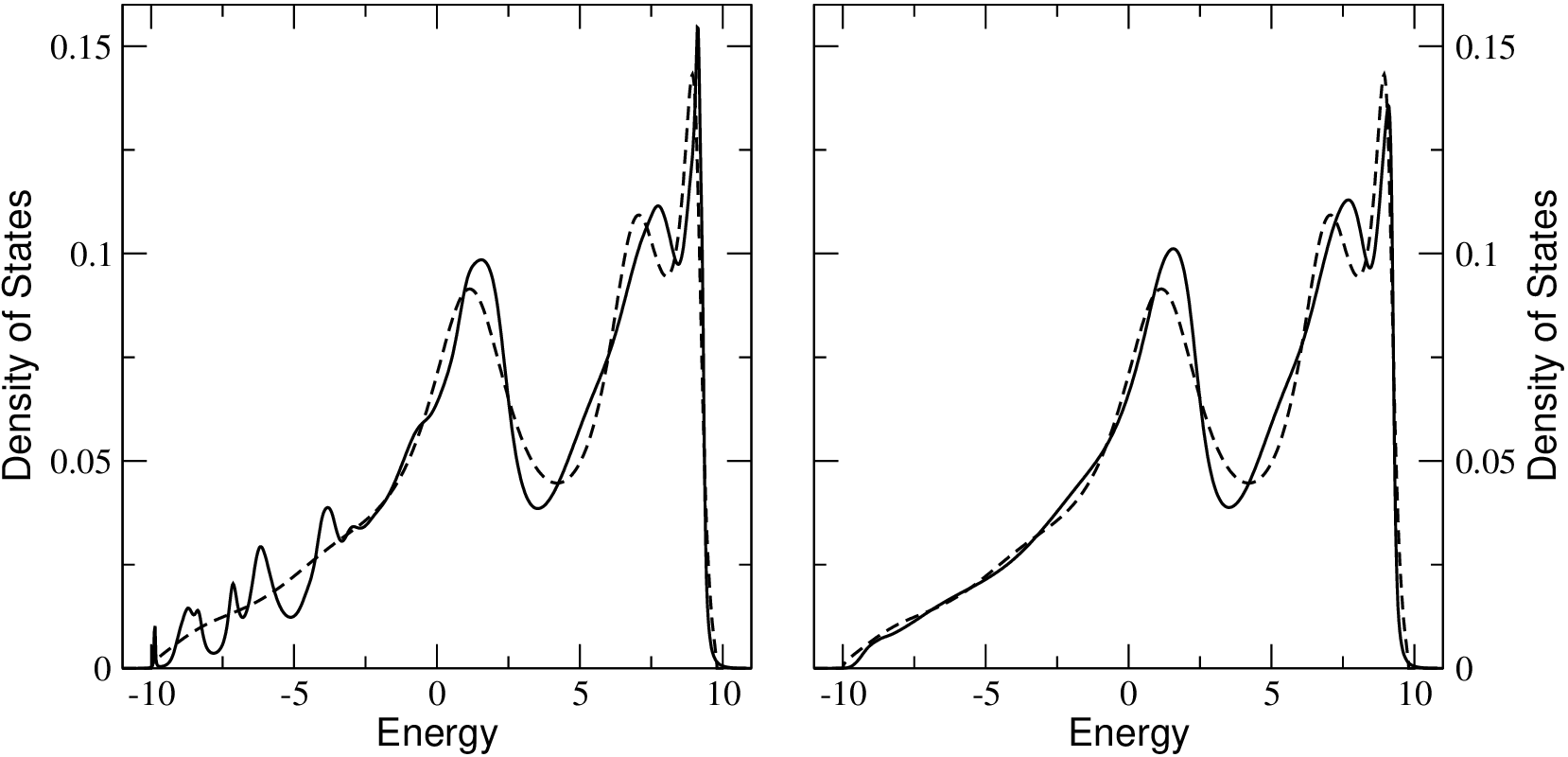}
\caption{\label{fig2}
Comparison of the average (50-50) density of states for a model fcc alloy calculated
using {\bf k-} space formulation of ASR (solid curve) with the same calculated using real-space 
formulation of ASR ((dotted curve). {\bf k-} space integration has been performed in two ways~: 
(a) using Tetrahedron Method (right solid curve) (b) multiplying spectral function $A_k(E)$ by k-point 
weight and then summing up over k (left solid curve). In both figures we note that the 
oscillations shown by the brute force technique is smoothened by the TM.}
\end{center}
\end{figure*}

We have first carried out calculations on a simple model disordered binary alloy system 
described by a $s$-state tight-binding Hamiltonian with nearest neighbour 
hopping integrals only. 
In Fig.~\ref{fig2} we compare the  results obtained using reciprocal and real
space formulation of ASR. The {\bf k-} space integration has been performed in two ways.
The brute force method, where we replace the integral by a sum with appropriate weights
at different $\k$-points, generates some unusual oscillations particularly in  the
lower part of the band. However,  the tetrahedron method gives 
smoother results which are in  good agreement with the real space calculations as well.

We now go over to calculations for the disordered Ni$_{50}$Pt$_{50}$ alloy. We have
used the minimal basis set of the TB-LMTO with nine orbitals per atom ($s$, $p$ and $d$)
to set up our Hamiltonian.

\begin{figure}[b]
\begin{center}
\includegraphics[width=3.2in, height=2.in]{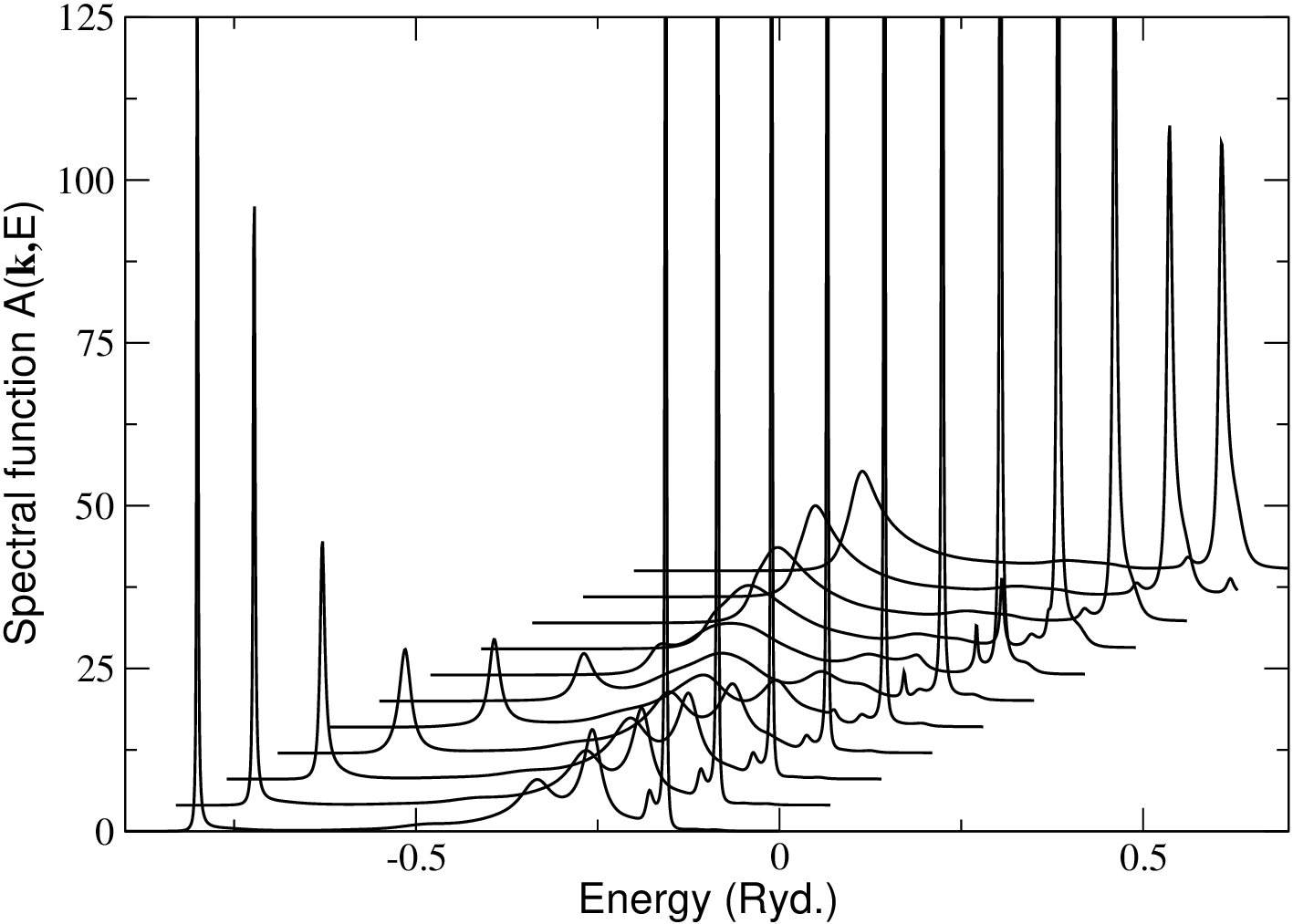}
\caption{\label {fig3}
The spectral function of Ni$_{50}$Pt$_{50}$ alloy plotted as a function of 
energy at several $\k$-points along the $\Gamma-X$ direction.}
\end{center}
\end{figure}

In Fig.~\ref{fig3} we present the results for the spectral functions for Ni$_{50}$Pt$_{50}$
alloy along the $\Gamma-X$ direction. We have chosen 11 k-points having equidistant between 
$\Gamma$ to $X$ points and have shown spectral function in those points. These spectral 
functions shows good agreement with the same results obtained from KKR-CPA calculations 
\cite{Gyorffy}. If we look carefully, we see that the widths of the spectral function
varies considerably as a function of $\k$ and $E$. There are some simple trends concerning
this behaviour. The sharp peaks on the lower band edge of near the $\Gamma$ 
point appear as the $s$-like band. As we go towards $\Gamma$ to $X$ point the $s$-band 
hybridizes with the $p$-band and the peak becomes wider. The structures on the upper
band edges are mostly due to the overlap of the $d$-states of Ni and Pt. Disorder 
affects to these $d$-dominated states are  strong and there is significant broadening. 

\begin{figure*}
\begin{center}
\includegraphics[width=2.5in, height=3.5in]{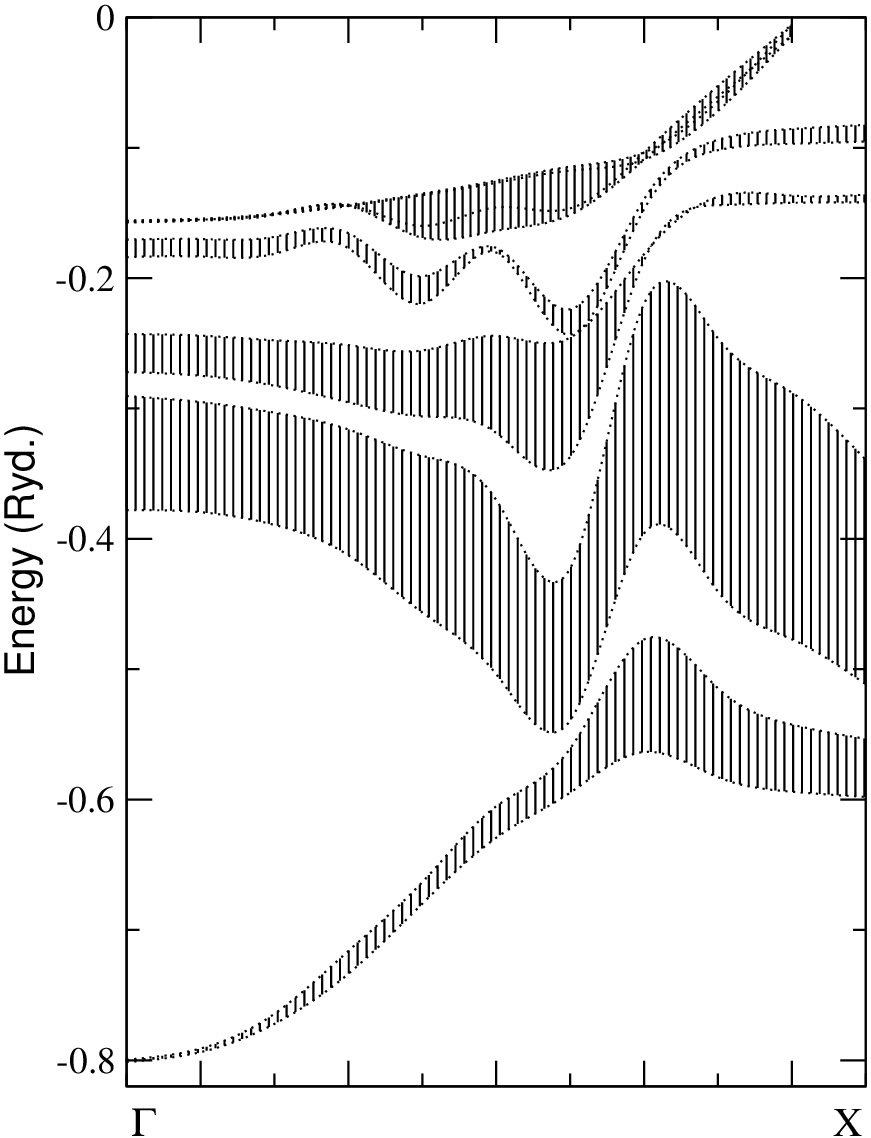}
\hskip 0.8cm
\includegraphics[width=2.5in, height=3.5in]{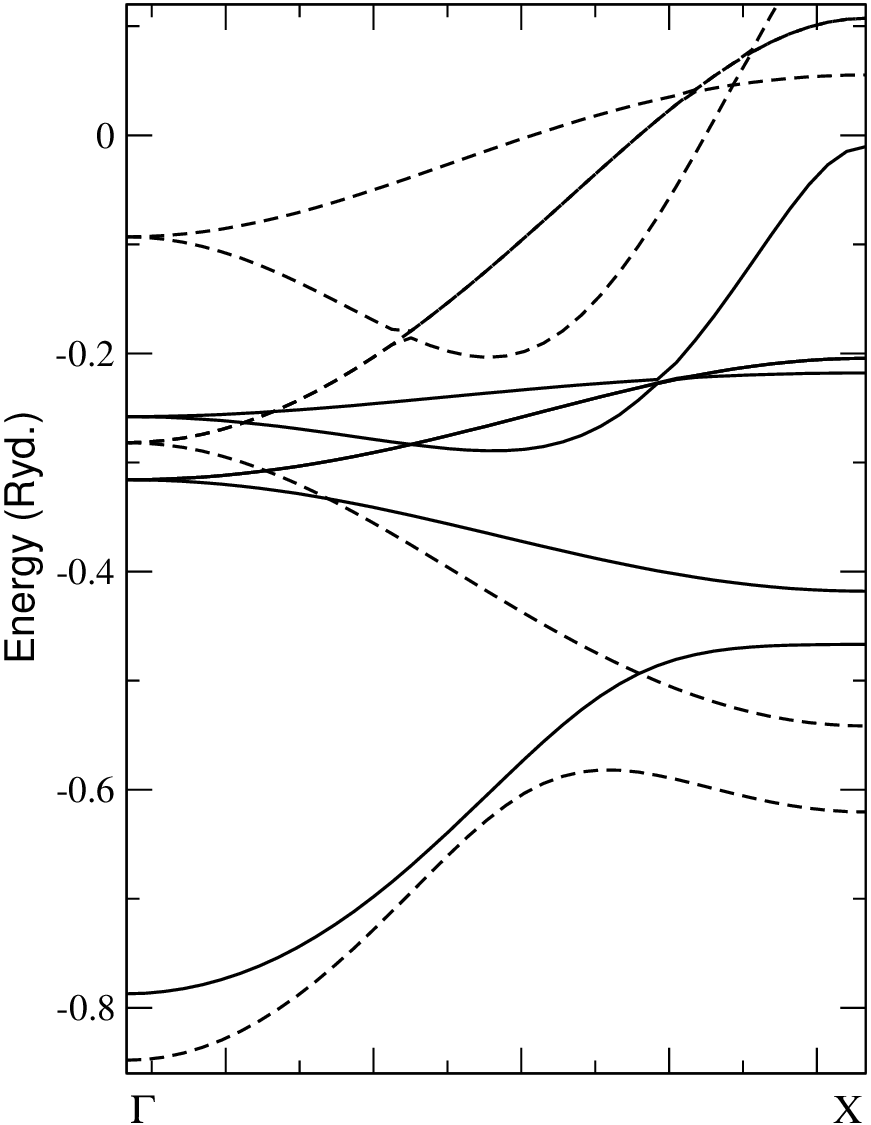}
\caption{\label {fig4}
(Right) Ni and Pt energy bands on a lattice appropriate to the Ni$_{50}$Pt$_{50}$ alloy, 
in the $\Gamma-X$ direction. Average lattice parameter $a_0=7.127 au$ was fixed after 
minimizing the energy. (Left) The fuzzy band of the disordered Ni$_{50}$Pt$_{50}$ system plotted 
along the same direction.}
\end{center}
\end{figure*}

\begin{figure*}
\begin{center}
\includegraphics[width=5.5in, height=5.0in]{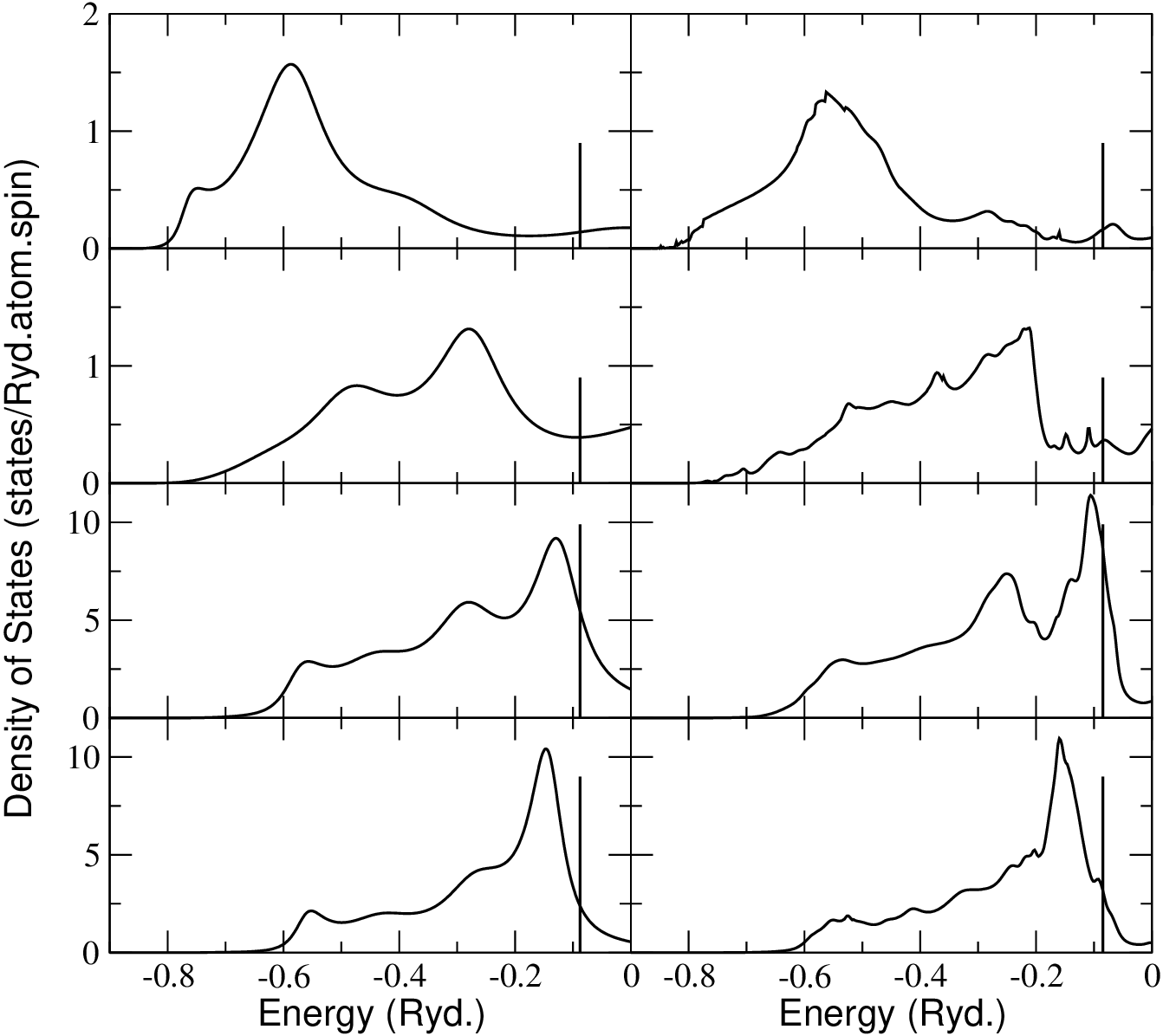}
\caption{\label{fig6}
Comparison of the partial density of states of Ni$_{50}$Pt$_{50}$ alloy calculated using 
augmented space recursion in (a) real-space formulation (left panel). (b) {\bf k-} space formulation (right panel).}
\end{center}
\end{figure*}

\begin{figure*}[t]
\begin{center}
\includegraphics[width=6.2in, height=4.2in]{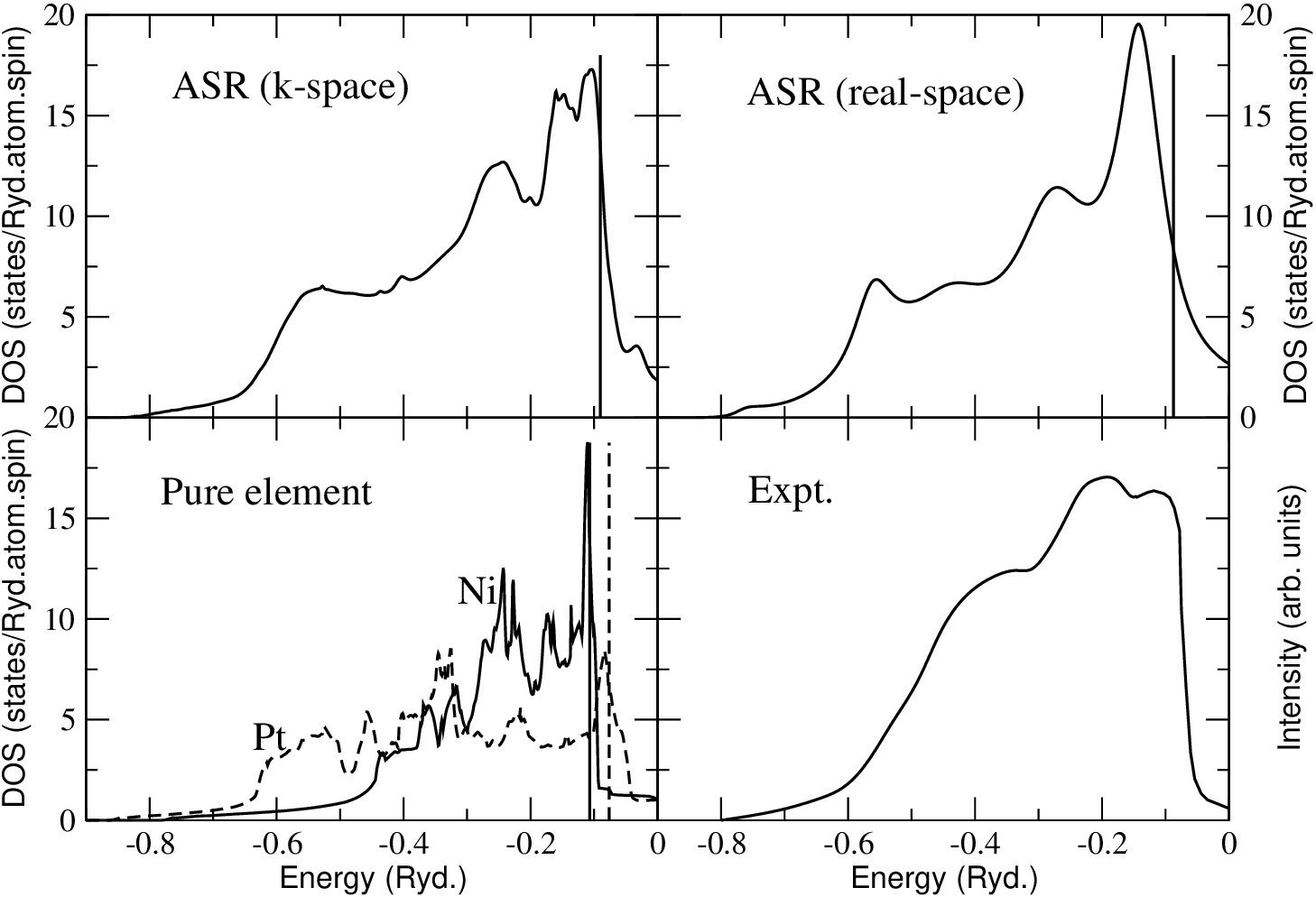}
\caption{\label{fig5}
Comparison of the density of states Ni$_{50}$Pt$_{50}$ alloy calculated using augmented space recursion
in (a) {\bf k-} space formulation. (b) real space formulation. (c) Density of states of Ni (solid line) and 
Pt (dotted line) on a lattice appropriate to the Ni$_{50}$Pt$_{50}$ alloy. 
(d) Valence-band photoemission spectra of Ni$_{50}$Pt$_{50}$ with photon energy $h\nu=60$.}
\end{center}
\end{figure*}

In Fig.~\ref{fig4} we presented the complex fuzzy bands of the disordered alloy.
The disorder smearing is maximum in the overlapping d-bands of the constituents, and is
negligible in the s-like part. This is also apparent in the spectral functions shown
earlier. The sharp s-like peaks flank wide d-like structures in Fig.~\ref{fig3}.

Finally using our modified tetrahedron method we have calculated the density of states (DOS) of ordered and homogeneous disordered NiPt alloys from its spectral function.
Side by side we have also carried out the same calculation in real augmented space.
In Fig.~\ref{fig6} we show the $\ell-$projected density of states for the
Ni$_{50}$Pt$_{50}$ alloy. We compare the k-space results with those found from
real space recursion. 
The main improvement occurs in the eg and t2g d-bands. In particular, the sharp
feature straddling the Fermi energy is better reproduced in the k-space recursion
than that in real space. The reason for this is the early truncation of recursion in real space and the consequent finite size effects to which the more localized
d-states are more susceptible.

In Fig.~\ref{fig5} (top row) we show a comparison
between average DOS calculated by real and reciprocal-space recursions. As 
discussed before, it is the sharp feature straddling the Fermi energy with major
contribution coming from the NMi d-states which are not well reproduced in the
real space technique. 
In this point our {\bf k-} space calculations agree
with the KKR-CPA results of  Staunton {\it et. al.} \cite{staunton}.
In the left lower panel of
Fig.~\ref{fig5} we show the DOS for pure Ni and Pt, but in a lattice with the lattice parameter same as
the alloy. We may compare this with the DOS for the disordered alloy. 
The right most three peaks at -0.25, -0.16 and -0.11 Ryd. of the
disordered DOS  are mostly contributed by  Ni whereas the left (lower energy) structures come 
(large shoulder at -0.57 Ryd.) mostly from Pt. 
The sharp peaks in the elemental results are obviously 
because of the Van Hove singularities of the DOS. The effect of disorder  
mainly smears out the sharp peaks present in the DOS 
. The disorder smearing is more pronounced for the d-like parts of the band.
 We remark that there is  very little 
shift in the DOS-related features between the ordered and disordered states.
 
Finally in the right lower panel we show the photoemmission spectrum of Ni$_{50}$Pt$_{50}$ reported by Nahm et.al. \cite{nahm}. The general features with a double peak
straddling the Fermi energy and a lower energy shoulder are clearly seen. The 
photoemmission spectra are convolutions of the density of states with a weakly
energy/wavenumber  dependent transition matrix. This may lead to shifting and smearing
of the prominent peak structures. Keeping this in mind, our k-space recursion
results are in good agreement with experiment.

\section{Remarks and Conclusion}

We have presented here an augmented space recursion formulation in reciprocal space. We also
present a generalization of the tetrahedron method proposed by Jepsen {\em et. al.} \cite{andersen} 
for inverting the spectral functions to obtain the density of states. This technique will be useful
for carrying of Brilluoin zone integrals for disordered alloys. We have studied both a 
model alloy and NiPt. The latter was so chosen as its has a sharp structure straddling the
Fermi energy and will be a sensitive test for the accuracy of our technique.

\section*{Acknowledgements} K. K. Saha would like to thank Profs. O. K. Andersen and Ove Jepsen
for their kind help in making the development of generalized {\bf k-} space integration for random 
alloys and their kind hospitality during his visit to the Max-Panck Institute, Stuttgart, Germany.

\appendix

\section{The Augmented Space Theorem}

Let $f(n_R)$ be a function of a random variable $n_R$, whose binary probability density is given
by~:
\[p(n_R) = x \ \delta(n_R-1) + y \ \delta(n_R) \]
We may then write~:

\[p(n_R)=-\frac{1}{\pi} \lim_{\delta\rightarrow 0} \ \Im m 
\langle \uparrow_R|\left((n_R+i\delta) {\bf I}-{\bf M}_R\right)^{-1}|\uparrow_R\rangle \]

Here, the operator ${\bf M}_R$ acts on a space spanned by the eigenvectors $|1_R\rangle$ and
$|0_R\rangle$ of ${\bf M}_R$, corresponding to eigenvalues 1 and 0; $|\uparrow_R\rangle=
\sqrt{x}|1_R\rangle+\sqrt{y}|0_R\rangle$ is called the {\sl reference state}. Its orthogonal
counterpart is $|\downarrow_R\rangle=\sqrt{y}|1_R\rangle-\sqrt{x}|0_R\rangle$. The representation
of ${\bf M}_R$ in this new basis~:
\[{\bf M}_R = \left(\begin{array}{cc}
x & \sqrt{x y} \\
\sqrt{x y} & y 
\end{array} \right) \]
Now,
\begin{eqnarray}
&&\ll f(n_R)\gg \ = \int_{-\infty}^\infty f(n_R) p(n_R) dn_R \nonumber \\
&&\qquad =-\frac{1}{\pi}\Im m\int_{-\infty}^\infty f(n_R) \langle\uparrow_R|{(n_R {\bf I}-{\bf M}_R)}^{-1}
|\uparrow_R\rangle \ dn_R \nonumber \\
&&\qquad = -\frac{1}{\pi}\Im m\sum_{\lambda=0,1}\sum_{\lambda'=0,1}f(n_R)\langle\uparrow_R|\lambda\rangle\dots\nonumber \\
&&\phantom{xxxxxx} \dots\langle\lambda|{(n_R{\bf I}-{\bf M}_R)}^{-1}|\lambda'\rangle\langle\lambda'|\uparrow_R\rangle \ dn_R \nonumber \\
&&\qquad = \sum_{\lambda=0,1}\langle\uparrow_R|\lambda\rangle f(\lambda)\langle\lambda|\uparrow_R\rangle\nonumber  \\ 
&&\qquad = \langle \uparrow_R|\mathbf{\tilde{f}}|\uparrow_R\rangle
\end{eqnarray}

Here $\mathbf{\tilde{f}}$ is an operator built out of $f(n_R)$ by simply replacing the variable
$n_R$ by the associated operator ${\bf M}_R$. The above expression shows that the average is
obtained by taking the matrix element of this operator between the {\sl reference state} 
$|\uparrow_R\rangle$. The full Augmented Space Theorem is a generalization of this for functions of many
independent random variables $\{n_R\}$. 

\section{Terminators}

The recursive calculation described earlier gives rise to a set of continued fraction 
coefficients $\{\alpha_n,\beta_n\}$. In any practical calculation we can go only upto a finite 
number of steps, consistent with our computational process. In case the coefficients converge, 
i.e. if $|\alpha_n-\alpha|\le \epsilon,$ $|\beta_n-\beta|\le \epsilon$ for $n\ge N$, we may 
replace $\{\alpha_n,\beta_n\}$ by $\{\alpha,\beta\}$ for all $n\ge N$. In that case the
asymptotic part of the continued fraction may be analytically summed to obtain~:

\[ \Gamma(E) = \frac{1}{2}\left(E-\alpha-\sqrt{{(E-\alpha)}^2-4\beta^2}\right) \]

which gives a continuous spectrum $\alpha-2\beta\ge E\ge\alpha+2\beta$. Since the terminator
coefficients are related to the band edges and widths, a sensible criterion for the choice of
these asymptotic coefficients is necessary, so as not to give arise to spurious structures in
our calculations. Beer and Pettifor \cite{bp} suggest a sensible criterion~: given a finite
number of coefficients, we must choose $\{\alpha,\beta\}$ in such a way so as to give, for
this set of coefficients, the minimum band width consistent with no loss of spectral weight
from the band. Let us call these values $\{\alpha_c,\beta_c\}$. This criterion is easily translated into mathematical terms. The delta function
that would carry weight out the band must then be situated exactly at the band edges. We thus
demand that the continued fraction diverge simultaneously at both the top and the bottom of
the band.

At the band edges~: $\Gamma(\alpha\pm2\beta)=\pm \beta$ and so,

\begin{widetext}
\[ \ll \!G(\alpha\pm 2\beta)\!\gg  = \! \frac{\beta_1^2/2}
        {\displaystyle \pm\beta-\frac{1}{2}(\alpha_1-\alpha)-\frac{\beta^2_2/4}
        {\displaystyle \pm\beta-\frac{1}{2}(\alpha_2-\alpha)-\frac{\beta^2_3/4}
        {\displaystyle \frac{\ddots \beta^2_N/2}
        {\displaystyle \pm\beta-(\alpha_N-\alpha)}}}}         \]
\end{widetext}

For a given $\alpha$, the $(N+1)$ eigenvalues of the finite tridiagonal matrix

\[ \left(\begin{array}{cccccc}
\frac{1}{2}(\alpha_1-\alpha) & \frac{1}{2}\beta_2 & 0 & \cdots & \cdots & 0 \\
\frac{1}{2}\beta_2 & \frac{1}{2}(\alpha_2-\alpha) & \frac{1}{2}\beta_3 & \ddots & \ & \vdots  \\
0 & \frac{1}{2}\beta_3 & \ddots & \ddots & \ddots & \vdots   \\
\vdots & \ddots & \ddots & \ddots & \ddots & 0 \\
\vdots & \ & \ddots & \ddots & \ddots & \frac{1}{\sqrt{2}}\beta_N \\
0 & \cdots & \cdots & 0 & \frac{1}{\sqrt{2}}\beta_N & (\alpha_N-\alpha)
\end{array} \right) \]

are values at which the Green function diverges. The maximum and minimum of this set of
eigenvalues are those values of $\beta$ for which spectral weight has just split off from
the band. Thus our choice of $\alpha$ is that value for which the maximum eigenvalue is
the largest and the minimum the smallest. Since the terminator only involves $\beta^2$ 
we must have

\[\beta_c = \sup_{\{\alpha\}} \ \beta_{max}(\alpha_c)=\inf_{\{\alpha\}} \ 
|\beta_{min}(\alpha_c)|  \]

With this choice the terminator $\Gamma(E)$ has all the Herglotz properties required.

\end{document}

%% file: macros.tex
\def\sumk{\int_{BZ} \frac{d^3\k}{8\pi^3}\ }
\def\o#1{\mathbf{#1}}

\def\ket{\vert \vert  \{ \emptyset \} \rangle}
\def\ket2{\vert \vert \otimes \{ R \} \rangle}

\def\n{\noindent }
\def\mbf#1{{\mathbf {#1}}}

\def\eq{\ =\ }

\def\be{\begin{equation}}
\def\ee{\end{equation}}
\def\<{[}
\def\>{]}

\def\k{{\mathbf k}}

%% file: paper.bbl
\begin{thebibliography}{99}
\bibitem{her} A function of a complex variable f(z) is called Herglotz if :
all singularities of f(z) lie on the real z axis and Im f(z)~=~$-$ sgn(Im z) for all
z off the real axis.
\end{thebibliography}
